# Anisotropic Li diffusion in pristine and defective ZnO

Ganes Shukri[1,2,*], Adhitya G. Saputro[1,2,*], Poetri S. Tarabunga[1], Febriyanti V. Panjaitan[1], Mohammad K. Agusta[1,2], Hermawan K. Dipojono[1,2]

[1]Advanced Functional Materials Research Group, Department of Engineering Physics, Faculty of Industrial Technology, Institut Teknologi Bandung, Bandung 40132, West Java, Indonesia

[2]Research Center for Nanoscience and Nanotechnology, Institut Teknologi Bandung, Bandung, 40132, Indonesia

*Corresponding Author: ganda@tf.itb.ac.id; ganes@tf.itb.ac.id

**Abstract**

We study the Li interstitial diffusion in pristine and defective ZnO bulk by means of first-principles density functional theory (DFT) coupled with Nudged Elastic Band (NEB) calculations. We consider three types of point defects, *i.e.*, oxygen vacancy ($O_{vac}$), Zn vacancy ($Zn_{vac}$) and ZnO vacancy pair ($ZnO_{vac\text{-}pair}$) and investigate their individual effect on the energy barrier of Li interstitial diffusion. Our results predict that $O_{vac}$ and $Zn_{vac}$ lower the Li diffusion energy barrier as compared to the pristine ZnO case. However, we further find that Li interstitial on the other hand may possibly be trapped inside the $Zn_{vac}$ subsequently forming the $Li_{Zn}$ substitutional type of defect. The similar behavior also observed for Li interstitial in the vicinity of Zn-$O_{vac\_pair}$ though with less change of Li diffusion barriers as compared to the other two cases. Our results indicate that among the three considered defects only $O_{vac}$ shows possible enhancement of the kinetics of Li diffusion inside ZnO bulk.

Keywords: ZnO, point defects, Li diffusion, Li-ion battery.

## I. Introduction

Artificial coating on the surface of Li-ion battery (LIB) cathode is an effective method to enhance the cathode stability at high operating voltage. Various materials such as binary metal oxides[1–4], phophates[5–7], fluorides[8–10] have been reported as potential protective coating materials for LIB cathode. Those previous studies have further shown that the enhanced stability of the cathode structure increase as well the capacity retention of LIB over prolonged charging-discharging cycles. The following mechanisms have been proposed as the physical origin of the aforementioned enhanced performance: (1) the coating shields the active cathode material from parasitic decomposition of the electrolyte on the cathode surface[11,12] (2) the coating scavenges HF molecule which can decrease local acidity at the cathode/electrolyte interface area, hence decreasing the electrolyte degradation[13,14] and (3) the coating mitigates structural deformation on the cathode surface, *e.g.*, surface oxygen loss, transition metal dissolution and surface cation migration.[15–19]

Based on previous studies there are two forms of coating in general. The first is nano particles coating which commonly covers the surface of the cathode partially. The second is thin film layer which encapsulates all of the cathode surface. In our present work, we will assume that all cathode area is covered by the thin film layer uniformly. In such scenario, direct contact between electrolyte and cathode surface which possibly lead to electrolyte decomposition as well as transition metal dissolution (i.e. corrosion) can be suppressed maximally. As a further consequence the coating ability to transport Li is likely to play an important role. Previous studies have investigated several important aspects that influence the Li transport inside the coating materials such as coating thickness, atomic structure, coating/cathode interfacial structure and defect chemistries[20–22]. The thickness of the coating layer is perhaps one of the most studied due to the easiness to control coating deposition relative to the other aforementioned aspects. For instance, by using Magnetron Sputtering Lai *et al* synthesized ZnO coated Ni-rich $Li(Ni_{0.8}Co_{0.15}Al_{0.05})O_2$ (NCA81505) cathode. They observed correlation between ZnO film thickness as a function of sputtering time with the electrochemical performance of the LIB. They found that the electrochemical performance increases when relatively thin ZnO film (~20 nm) was deposited onto the cathode surface but then reduces when the ZnO film became thicker.[23] Similar insight was also obtained by Riley *et al* in which Atomic Layer Deposition was used to grow an $Al_2O_3$ film on $Li(Ni_{0.33}Co_{0.33}Mn_{0.33})O_2$. They observed that Li transport is significantly hindered as the $Al_2O_3$ coating thickness exceeds ~12 A.[24]

While studies related to coating thickness and Li transport have been conducted intensively, study related to the role of defect on Li transport in coating materials in the context of LIBs application is relatively scarce. Herein, we focus on the ZnO one of the widely used surface protective material for Li-ion battery cathode. While previous studies have shown that ZnO indeed can increase the capacity retention of the overall Li-ion battery cycles[20,23,25–27] the quantitative knowledge related to how point defects affect the Li diffusion in ZnO remains unclear. Our results clarify that while some certain point defects hinder the Li diffusion inside ZnO, oxygen vacancy promotes faster Li diffusion as shown by relatively low diffusion energy barrier.

## II. Computational Setup

We carried out spin-polarized density functional theory (DFT) calculations as implemented in quantum ESPRESSO package.[28] Ultrasoft pseudopotentials were used to parametrize the core region of each elements used in the present work. The Generalized Gradient Approximation of Perdew-Burke-Ernzerhof (GGA-PBE) was used to describe the exchange interactions and electronic correlations.[29] We applied 42 Ry kinetic cut-off energy for the expansion of Kohn-Sham wave functions. Brillouin zone integrations for bulk geometry optimizations and electronic structure calculations were done with 4x4x4 *k*-points sampling. Based on these parameters, the calculated lattice constants for the unit cell of wurtzite ZnO bulk are $a$ = 3.26 Å and $c/a$ = 1.61 which are in good agreement with previous results.[30] Hubbard *U* term was included in the DFT Hamiltonian (hence DFT+*U*) to correct the self-interaction error on Zn 3*d* orbital. An effective *U* value of 7.5 eV was then applied to the Zn 3*d* orbital. The chosen *U* value has been successfully used to predict the correct electronic band gap of pristine ZnO bulk as previously reported.[31–34] For Li interstitial diffusion energy barrier calculation, we utilized the nudged-elastic band (NEB) method with four transition images between each initial and final configuration.

## III. Results and Discussion

Based on the ZnO structure, we considered two different hollow sites occupied by the interstitial Li (Li$_i$), *i.e.*, the octahedra and the tetrahedra sites. The octahedra site is a hollow site surrounded by six oxygen ions whereas the tetrahedra is surrounded by four oxygen ions. Geometry optimizations show that Li$_i$ located on the octahedra site (Li$_{i\text{-}oct}$) is thermodynamically more stable as compared to the tetrahedra site (Li$_{i\text{-}tet}$) by 1.04 eV. We also find that further geometry optimization if preceded by a slight distortion, will move the Li$_{i\text{-}tet}$ to the octahedra site (hence becoming Li$_{i\text{-}oct}$). This finding indicates that in general octahedra is the most preferred site for Li$_{i\text{-}oct}$ residing within pristine ZnO. Note that for most of the subsequent initial Li interstitial diffusion calculation in ZnO with defect, we then used the Li$_{i\text{-}oct}$ as the initial configuration unless otherwise stated.

For the case of Li diffusion in pristine ZnO, we identified two possible diffusion pathways: (i) parallel to the (001) basal plane (or octahedra-to-tetrahedra hence 'O-T') path, and (ii) along the [001] direction (or direct edge-sharing octahedra-to-octahedra hence 'O-O') path (**Figure 1a and 1b**). Our calculated Li$_i$ diffusion barrier (E$_{diff\_bar}$) in pristine ZnO show that the O-T path has a lower E$_{diff\_bar}$ as compared to the O-O path (1.05 eV vs 1.25 eV, respectively). Note that however, as mentioned before, Li$_{i\text{-}tet}$ is less stable than the Li$_{i\text{-}oct}$. As can be seen in **Figure 1b**, diffusion of Li$_i$ from tetrahedra to octahedra site needs only a small energy barrier (~0.01 eV). Hence, Li$_{i\text{-}tet}$ most probably will migrate to its neighboring octahedra site, effectively forming the octahedra-tetrahedra-octahedra (hence, 'O-T-O') diffusion path. Our calculated E$_{diff\_bar}$ are in good agreement with the previous experimental values of interstitial Li$_i^+$ diffusion in wurtzite ZnO.[35,36] Though, indeed we also found some discrepancies of our calculated Li$_i$ diffusion energies with previous theoretical works.[30,37] We argue that the discrepancies are due to the difference of exchange-correlation functional used in our present work with the previous theoretical calculations, *i.e.*, PBE+$U$ (in our present work) and LSDA (previous works). Nevertheless, we note that the *qualitative* finding is in good agreement in which the 'O-T-O' is the energetically more preferred path for Li$_i$ diffusion in the pristine ZnO case.

To determine the effect of point defects on the energetics of Li$_i$ diffusion, we considered three types of point defect, *i.e.*, oxygen vacancy (O$_{vac}$), Zn vacancy (Zn$_{vac}$) and Zn-O vacancies pair (Zn-O$_{vac\_pair}$). Our calculated formation energies for each point defects are $E^f_{O_{vac}}$ = 4.51 eV, $E^f_{Zn_{vac}}$ = 6.22 eV and $E^f_{Zn-O_{vac\_pair}}$ = 7.3 eV, respectively, which are in reasonable agreement with previous theoretical results.[38]

Next, we investigated the effect of O$_{vac}$ on the energetics of Li$_i$ diffusion. Note that here we only considered the formation of neutral O$_{vac}$, hence two $e^-$ remain in the ZnO upon the formation of the O$_{vac}$.[38,39] We then evaluated the most stable configuration of Li$_i$ in the vicinity of the O$_{vac}$. Thermodynamically, Li$_i$ tends to occupy the octahedral site located further from the O$_{vac}$ site, *i.e.*, the next-nearest and next-next-nearest octahedral from O$_{vac}$ (as can be inferred from the initial-final position of Li$_{i\text{-}oct}$ diffusion of **Figure 2a-b**). This is so in order to minimize the Coulomb repulsion between the Li$_i$ and O$_{vac}$ as each of both defects can be thought as an $e^-$ donor to the ZnO. Calculation of Li$_i$ diffusion barriers through O-T and O-O paths with initial configuration of Li$_{i\text{-}oct}$ at next-nearest octahedra to the O$_{vac}$ show significant decreased diffusion energy barriers (0.82 eV and 0.26 eV, respectively) as compared to the pristine ZnO case (**Figure 2a-b**). Note as well that although at first, we considered the O-T diffusion path, we find that upon geometry optimization Li$_{i\text{-}oct}$ is more favorable compared to the initially considered Li$_{i\text{-}tet}$ as the final configuration. Hence, effectively the O-T-O is most likely the preferred diffusion path in the area close to the O$_{vac}$. Our results indicate that Li$_{i\text{-}oct}$ diffusion is kinetically faster in the vicinity of O$_{vac}$ and is thermodynamically favoring the diffusion direction away from the O$_{vac}$. We note as well that Li$_i$ diffusions starting from the next-next-nearest octahedra relative from the O$_{vac}$ site give diffusion energy barriers of 0.98 eV and 1.38 eV, respectively, which are close to the pristine case (**Figure 2c-d**). It can be seen from the results that $Li_i$ diffusion located further from the $O_{vac}$ tends to follow the behavior of that the pristine case.

For the case of ZnO with $Zn_{vac}$ we found that $Li_i$ tends to occupy the $Zn_{vac}$ site (hence becoming $Li_{Zn}$). In contrast to the $O_{vac}$, neutral $Zn_{vac}$ acts as electron acceptor whereas $Li_i$ acts as an $e^-$ donor thus resulting in attractive interaction. Calculated dissociation energy of $Li_{Zn}$ (*i.e.*, $Li_{Zn}$ → $Li_i$ + $Zn_{vac}$) shows that 3.98 eV of energy is needed for the reaction to occur. This finding is consistent with previous theoretical study which reported that Li is thermodynamically favorable as a substitutional dopant for Zn in ZnO [37]. However, this further implies that $Zn_{vac}$ may trap the $Li_i$ which can further hinder the $Li_i$ mobility. As can be seen in **Figure 3a-b**, $Li_i$ tends to move towards the $Zn_{vac}$ site shown by the enthalpy migration of the initial-final configurations for both O-O and O-T path (-0.18 eV and -1.23 eV, respectively). Calculated $Li_i$ diffusion barrier of O-O and O-T paths close to the $Zn_{vac}$ show energy barriers of 1.10 eV and 0.08 eV, respectively (**Figure 3a-b**). These indicate that, $Li_i$ has a faster mobility while moving on the basal plane around the $Zn_{vac}$ though with the possibility to be trapped inside the $Zn_{vac}$ itself. As we calculated $Li_i$ diffusion barrier with initial configuration of next-next-nearest neighbor for both O-T and O-O paths, the calculated $Li_i$ diffusion barriers remain close to the pristine ZnO case (1.13 eV and 1.21 eV, respectively) (**Figure 3c-d**). These indicates that only in the direct vicinity of $Zn_{vac}$ that the $Li_i$ diffusivity can be affected by the defect.

Next, we discuss the $Li_i$ diffusion in the vicinity of Zn-O vacancies defect pair ($Zn-O_{vac\_pair}$). It is interesting to note that both vacancies considered in this work tend to form an interacting $Zn_{vac}$ – $O_{vac}$ complex vacancies pair compared to two distinct isolated vacancy defects. This is indicated by the relatively stable binding energy, *i.e.*, 3.42 eV. We also found that similar to the isolated $Zn_{vac}$ defect case, $Li_i$ tends to occupy the $ZnO_{vac\_pair}$ forming the $Li_{Zn}$-$O_{vac}$ complex. The complex is thermodynamically stable with dissociative energy of $Li_{Zn}$ → $Li_i$ + $Zn-O_{vac\_pair}$ is calculated to be 3.56 eV. We then calculated the $Li_i$ diffusion at O-O path *close to* and *far from* the defect pair with the starting configuration of $Li_{i-oct}$ at next-nearest and next-next-nearest octahedral sites. Note that here, due to the cell size that we used in the present work, only the O-O diffusion path that was considered for the $Zn-O_{vac\_pair}$ case. We then found the $E_{diff\_bar}$ to be 1.01 eV and 1.40 eV, respectively (**Figure 4a-b**). While kinetic-wise the $Zn-O_{vac\_pair}$ may give the same tendency with the $Li_i$ diffusion in the pristine case the thermodynamic factor could be the difference due to the $Li_i$ tendency to occupy the $Zn-O_{vac\_pair}$ site. Hence, this could give similar behavior to that of $Zn_{vac}$ case where $Li_i$ could possibly trapped inside the defect site that may further hinder its mobility.

In order to compare our predicted $Li_i$ diffusion energy barriers in pristine and defective ZnO with respect to a condition similar to a LIB operating condition, we estimated the Li migration energy based on the transition state theory (TST). Here, we adopted the approach done by Xu et al.[20] Their model relates the film thickness, Li solubility and Li diffusion with overpotential at a given current. The model can further provide a qualitative guideline for what coating properties are necessary to maintain acceptably low losses in the battery. According to the TST, the ionic diffusivity can be estimated by the following Arrhenius equation:

$$D = D_0 \exp\left(-\frac{E_a}{k_B T}\right) \quad (1).$$

Herein, $E_a$ is the ion (Li) diffusion energy barrier, $k_B$ is the Boltzmann constant and $T$ is the estimated operating temperature. The $D_0$ is taken from previous experimental value of Li ion diffusion in ZnO, i.e., 0.02 cm²/s.[35] Note that for the case of LIB, the Li ion conduction is driven by the potential difference between the positive and negative electrodes. Hence, the Li conductivity can be evaluated according to the following relation,[20]

$$\sigma = Cq\mu = \frac{q^2 CD}{k_B T} \quad (2).$$

Here, $C$ is the ion concentration and $D$ can be stated by using Einstein relation for charged particle:

$$D = \frac{k_B T \mu}{q} \quad (3).$$

Next, the relation between overpotential of a solid film with a thickness $L$ and current density $J$ can be evaluated based on the following relation:

$$\Delta V = \frac{JL}{q\mu C} = \frac{JL k_B T}{DCq^2} \quad (4).$$

By inserting the value of $D_0 = 0.02$ cm²/s, $L = 10$ nm and some typical values of LIBs operating condition, i.e., $J = 0.046$ mA/cm², $T = 300$ K and $C = 10^{23}$ cm⁻³, we obtained the $Li^+$ activation energy to be ~0.80 eV. Comparing our DFT-predicted Li diffusion energies to the obtained activation energy indicates that $Li_i$ diffusion in ZnO is reasonably facile with respect to the typical operating condition of LIB should be. The introduction of $O_{vac}$ (or in general de-coordination of Zn-O bond) can possibly enhance the $Li_i$ mobility as previously explained. As for the $Zn_{vac}$ defect, although it lowers as well the $Li_i$ diffusion energy moving on the basal plane, it has a tendency to trap the $Li_i$ inside itself (similarl to the $Zn-O_{vac\_pair}$ case). Hence, up to this end, we conclude that *only* the $O_{vac}$ that can possibly benefit the $Li_i$ diffusion inside ZnO.

## IV. Conclusion

By means of first-principles density functional theory (DFT) calculations coupled with nudged elastic band (NEB) method we have investigated the effect of two point and one complex defects on $Li_i$ diffusion barrier inside ZnO bulk. Our results show that $O_{vac}$ lower the $Li_i$ diffusion energy barrier on both O-O and O-T(-O) paths as compared to the pristine ZnO case. For the case of $Zn_{vac}$, the $Li_i$ mobility also can increase as shown by the lowered diffusion barrier on O-T diffusion path but the with tendency of $Li_i$ moving *toward* the defect site. While kinetically it may increase $Li_i$ mobility, the formation of $Li_{Zn}$ may thermodynamically prohibits the overall $Li_i$ diffusion due to the $Li_i$ being "trapped" in the $Zn_{vac}$ site. This behavior also similarly observed for $Zn-O_{vac\_pair}$ case. To this end, the present findings imply that among the three defects considered, only $O_{vac}$ that can possibly enhance Li interstitial diffusion inside ZnO. We suggest that synthesizing reduced ZnO with $O_{vac}$ defect can be beneficial for Li mobility inside ZnO.

**Acknowledgement**

GS and AGS would like to acknowledge the financial support from "P3MI ITB 2020".

**References**

1. Kong, J. Z. *et al.* Enhanced electrochemical performance of LiNi0.5Co0.2Mn0.3O2 cathode material by ultrathin ZrO2 coating. *J. Alloys Compd.* **657**, 593–600 (2016).
2. Ren, T. *et al.* Enhancing the high-voltage performances of Ni-rich cathode materials by homogeneous La2O3 coating via a freeze-drying assisted method. *Ceram. Int.* **44**, 14660–14666 (2018).
3. Yao, J., Wu, F., Qiu, X., Li, N. & Su, Y. Effect of CeO2-coating on the electrochemical performances of LiFePO4/C cathode material. *Electrochim. Acta* **56**, 5587–5592 (2011).
4. Guo, S. *et al.* Surface coating of lithium-manganese-rich layered oxides with delaminated MnO2 nanosheets as cathode materials for Li-ion batteries. *J. Mater. Chem. A* **2**, 4422–4428 (2014).
5. Lu, Y. C., Mansour, A. N., Yabuuchi, N. & Shao-Horn, Y. Probing the origin of enhanced stability of AlPO4 nanoparticle coated liCoO2 during cycling to high voltages: Combined XRD and XPS studies. *Chem. Mater.* **21**, 4408–4424 (2009).

6. Cho, J. *et al.* Comparison of Al2O3- and AlPO4-coated LiCoO2 cathode materials for a Li-ion cell. *J. Power Sources* **146**, 58–64 (2005).

7. Li, G., Yang, Z. & Yang, W. Effect of FePO4 coating on electrochemical and safety performance of LiCoCO2 as cathode material for Li-ion batteries. *J. Power Sources* **183**, 741–748 (2008).

8. Xu, K. *et al.* Synthesis and electrochemical properties of CaF 2-coated for long-cycling Li[Mn 1/3Co 1/3Ni 1/3]O 2 cathode materials. *Electrochim. Acta* **60**, 130–133 (2012).

9. Sun, Y. K. *et al.* The role of AlF 3 coatings in improving electrochemical cycling of Li-enriched nickel-manganese oxide electrodes for Li-ion batteries. *Adv. Mater.* **24**, 1192–1196 (2012).

10. Lee, K. S., Myung, S. T., Kim, D. W. & Sun, Y. K. AlF3-coated LiCoO2 and Li[Ni1/3Co 1/3Mn1/3]O2 blend composite cathode for lithium ion batteries. *J. Power Sources* **196**, 6974–6977 (2011).

11. Chen, C. *et al.* High-performance lithium ion batteries using SiO2-coated LiNi0.5Co0.2Mn0.3O2 microspheres as cathodes. *J. Alloys Compd.* **709**, 708–716 (2017).

12. Song, G., Zhong, H., Dai, Y., Zhou, X. & Yang, J. WO 3 membrane-encapsulated layered LiNi 0.6 Co 0.2 Mn 0.2 O 2 cathode material for advanced Li-ion batteries. *Ceram. Int.* **45**, 6774–6781 (2019).

13. Li, X. *et al.* Significant impact on cathode performance of lithium-ion batteries by precisely controlled metal oxide nanocoatings via atomic layer deposition. *J. Power Sources* **247**, 57–69 (2014).

14. Aykol, M., Kirklin, S. & Wolverton, C. Thermodynamic aspects of cathode coatings for lithium-ion batteries. *Adv. Energy Mater.* **4**, 1–11 (2014).

15. Liu, T., Zhao, S. X., Wang, K. & Nan, C. W. CuO-coated Li[Ni0.5Co0.2Mn0.3]O 2 cathode material with improved cycling performance at high rates. *Electrochim. Acta* **85**, 605–611 (2012).

16. Li, XuLi, X. et al. (2014) 'Spinel LiNi0.5Mn1.5O4 as superior electrode materials for lithium-ion batteries: Ionic liquid assisted synthesis and the effect of CuO coating', Electrochimica Acta. Elsevier Ltd, 116, pp. 278–283. doi: 10. 1016/j. electacta. 2013. 11. 055. elian., Guo, W., Liu, Y., He, W. & Xiao, Z. Spinel LiNi0.5Mn1.5O4 as superior electrode materials for lithium-ion batteries: Ionic liquid assisted synthesis and the effect of CuO coating. *Electrochim. Acta* **116**, 278–283 (2014).

17. Dai, S. *et al.* Ultrathin-Y2O3-coated LiNi0.8Co0.1Mn0.1O2 as cathode materials for Li-ion batteries: Synthesis, performance and reversibility. *Ceram. Int.* **45**, 674–680 (2019).

18. Chang, W., Choi, J. W., Im, J. C. & Lee, J. K. Effects of ZnO coating on electrochemical performance and thermal stability of LiCoO2 as cathode material for lithium-ion batteries. *J. Power Sources* **195**, 320–326 (2010).

19. Shi, Y., Zhang, M., Qian, D. & Meng, Y. S. Ultrathin Al2O3 Coatings for Improved Cycling Performance and Thermal Stability of LiNi0.5Co0.2Mn0.3O2 Cathode Material. *Electrochim. Acta* **203**, 154–161 (2016).

20. Xu, S. *et al.* Lithium transport through lithium-ion battery cathode coatings. *J. Mater.*

*Chem. A* **3**, 17248–17272 (2015).

21. Bettge, M. *et al.* Improving high-capacity Li1.2Ni0.15Mn 0.55Co0.1O2-based lithium-ion cells by modifiying the positive electrode with alumina. *J. Power Sources* **233**, 346–357 (2013).
22. Kong, J. Z. *et al.* Improved electrochemical performance of Li1.2Mn0.54Ni0.13Co0.13O2cathode material coated with ultrathin ZnO. *J. Alloys Compd.* **694**, 848–856 (2017).
23. Lai, Y. Q. *et al.* Optimized structure stability and electrochemical performance of LiNi0.8Co0.15Al0.05O2 by sputtering nanoscale ZnO film. *J. Power Sources* **309**, 20–26 (2016).
24. Riley, L. A. *et al.* Electrochemical effects of ALD surface modification on combustion synthesized LiNi1/3Mn1/3Co1/3O2 as a layered-cathode material. *J. Power Sources* **196**, 3317–3324 (2011).
25. Yu, R., Lin, Y. & Huang, Z. Investigation on the enhanced electrochemical performances of Li1.2Ni0.13Co0.13Mn0.54O2 by surface modification with ZnO. *Electrochim. Acta* **173**, 515–522 (2015).
26. Xie, J. *et al.* Determination of Li-ion diffusion coefficient in amorphous Zn and ZnO thin films prepared by radio frequency magnetron sputtering. *Thin Solid Films* **519**, 3373–3377 (2011).
27. Arrebola, J. C., Caballero, A., Hernán, L. & Morales, J. Re-examining the effect of ZnO on nanosized 5 V LiNi0.5Mn1.5O4 spinel: An effective procedure for enhancing its rate capability at room and high temperatures. *J. Power Sources* **195**, 4278–4284 (2010).
28. Giannozzi, P. *et al.* QUANTUM ESPRESSO: A modular and open-source software project for quantum simulations of materials. *J. Phys. Condens. Matter* **21**, (2009).
29. Perdew, J. P., Burke, K. & Ernzerhof, M. Generalized gradient approximation made simple. *Phys. Rev. Lett.* **77**, 3865–3868 (1996).
30. Carvalho, A., Alkauskas, A., Pasquarello, A., Tagantsev, A. K. & Setter, N. A hybrid density functional study of lithium in ZnO: Stability, ionization levels, and diffusion. *Phys. Rev. B - Condens. Matter Mater. Phys.* **80**, 1–12 (2009).
31. Saputro, A. G. *et al.* Effect of surface defects on the interaction of the oxygen molecule with the ZnO(1010) surface. *New J. Chem.* **44**, 7376–7385 (2020).
32. Septiani, N. L. W. *et al.* Hollow Zinc Oxide Microsphere-Multiwalled Carbon Nanotube Composites for Selective Detection of Sulfur Dioxide. *ACS Appl. Nano Mater.* **3**, 8982–8996 (2020).
33. Fitriana *et al.* Enhanced NO Gas Performance of (002)-Oriented Zinc Oxide Nanostructure Thin Films. *IEEE Access* **7**, 155446–155454 (2019).
34. Saputro, A. G. *et al.* www.jmolekul.com. (2019).
35. Lander, J. J. Reactions of Lithium as a donor and an acceptor in ZnO. *J. Phys. Chem. Solids* **15**, 324–334 (1960).
36. Knutsen, K. E., Johansen, K. M., Neuvonen, P. T., Svensson, B. G. & Kuznetsov, A. Y. Diffusion and configuration of Li in ZnO. *J. Appl. Phys.* **113**, (2013).
37. Wardle, M. G., Goss, J. P. & Briddon, P. R. Theory of Li in ZnO: A limitation for Li-based p-type doping. *Phys. Rev. B - Condens. Matter Mater. Phys.* **71**, 1–10 (2005).

38. Vidya, R. *et al.* Energetics of intrinsic defects and their complexes in ZnO investigated by density functional calculations. *Phys. Rev. B - Condens. Matter Mater. Phys.* **83**, 1–12 (2011).
39. Shukri, G., Diño, W. A., Dipojono, H. K., Agusta, M. K. & Kasai, H. Enhanced molecular adsorption of ethylene on reduced anatase TiO2 (001): Role of surface O-vacancies. *RSC Adv.* **6**, 92241–92251 (2016).

**Figures**

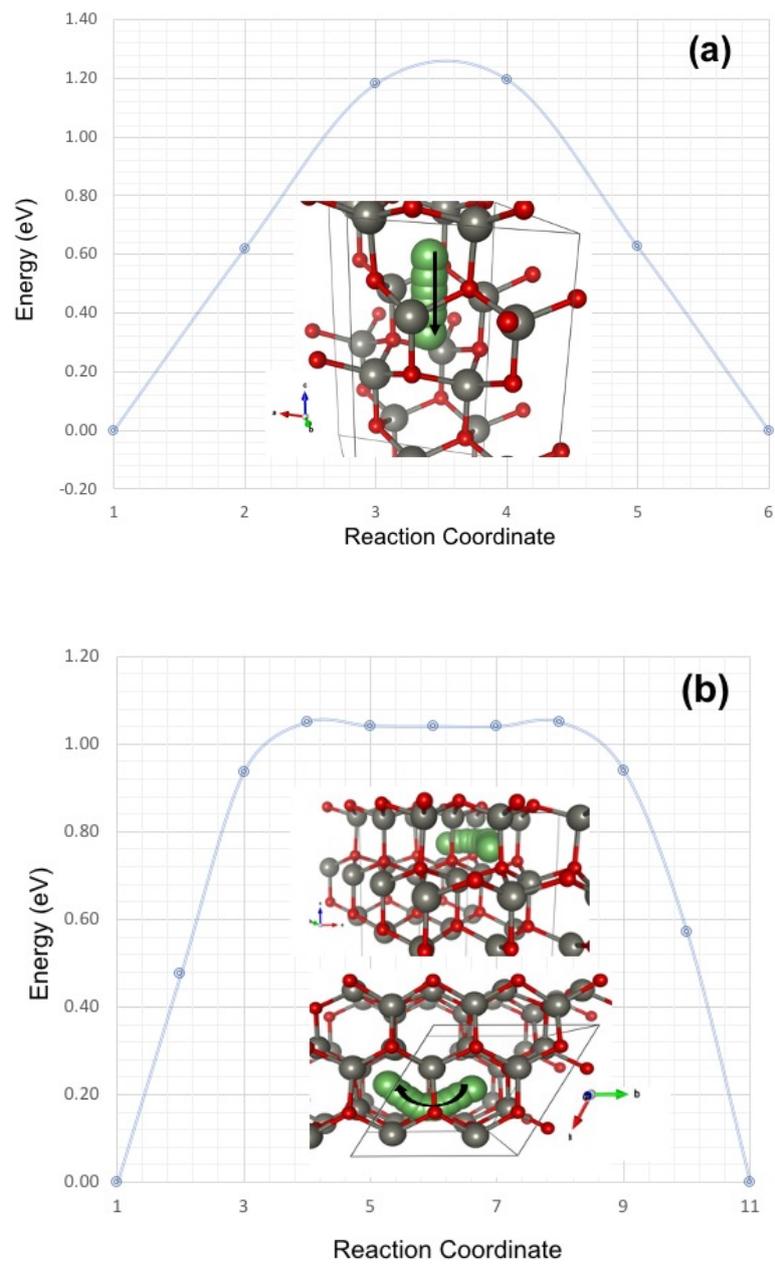

**Figure 1**. Li diffusion on pristine ZnO. (a) O-O path (side view) and (b) O-T-O path (side and top views).

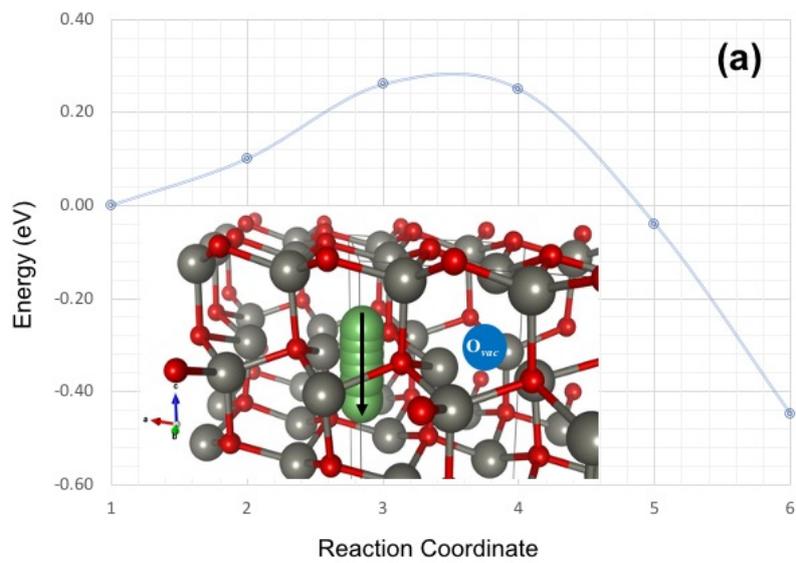

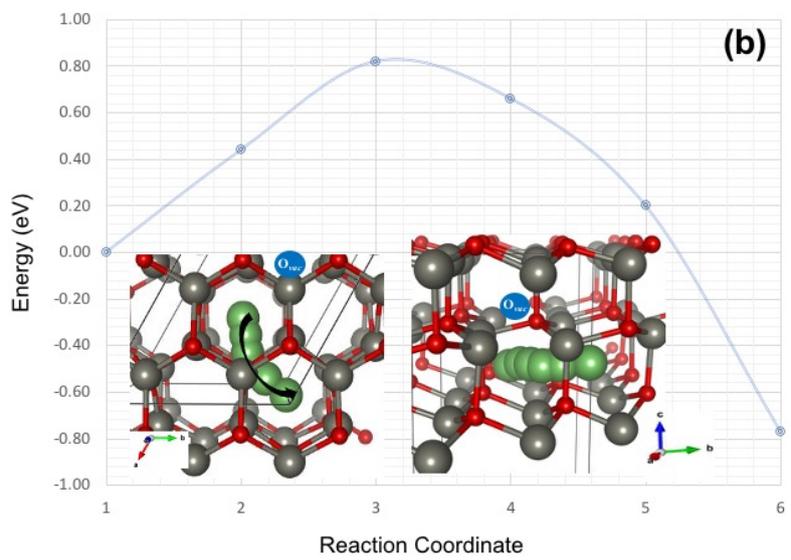

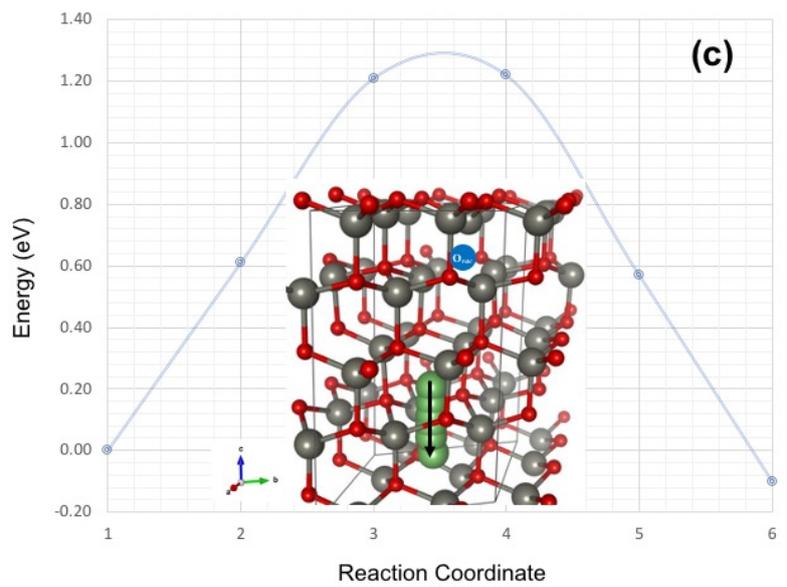

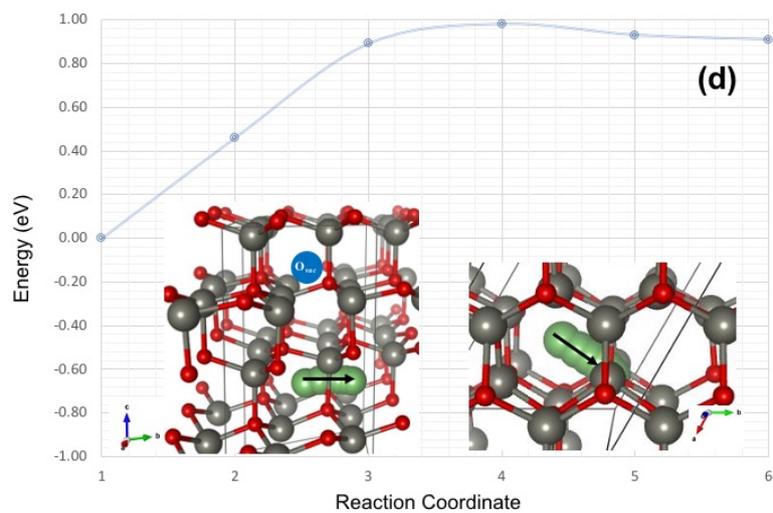

**Figure 2**. Li diffusion on reduced ZnO. (a) O-O path (side view) and (b) O-T Li diffusion paths (top and side views) at next-nearest octahedra relative from the $O_{vac}$. (c) O-O path (side view) and (d) O-T diffusion path (side and top views) at next-next-nearest octahedra relative from $O_{vac}$ site. $O_{vac}$ is marked by the blue circle.

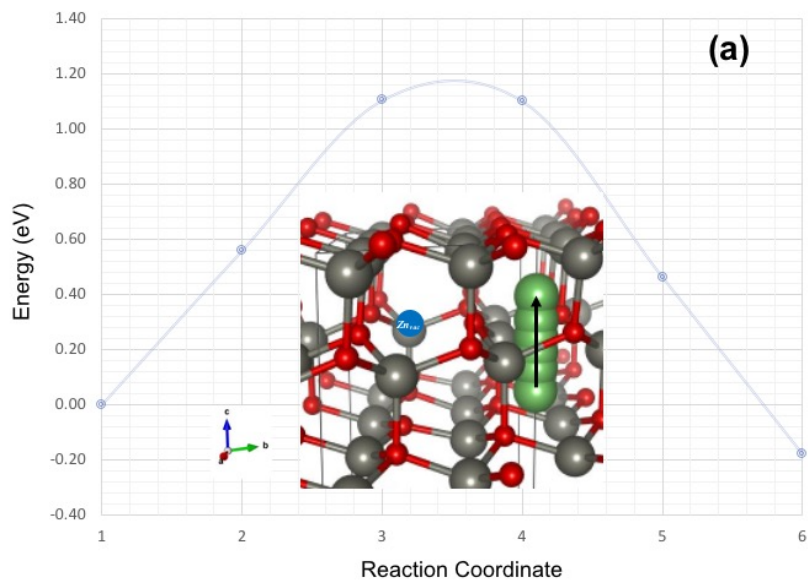

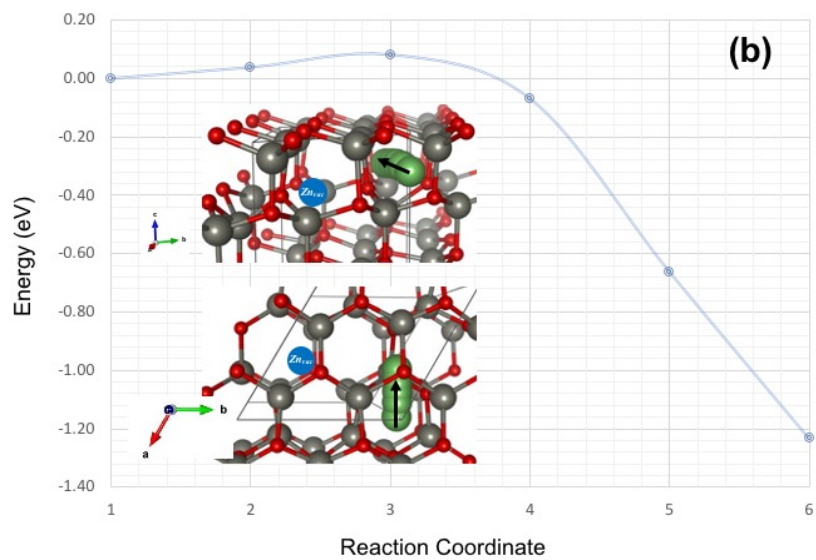

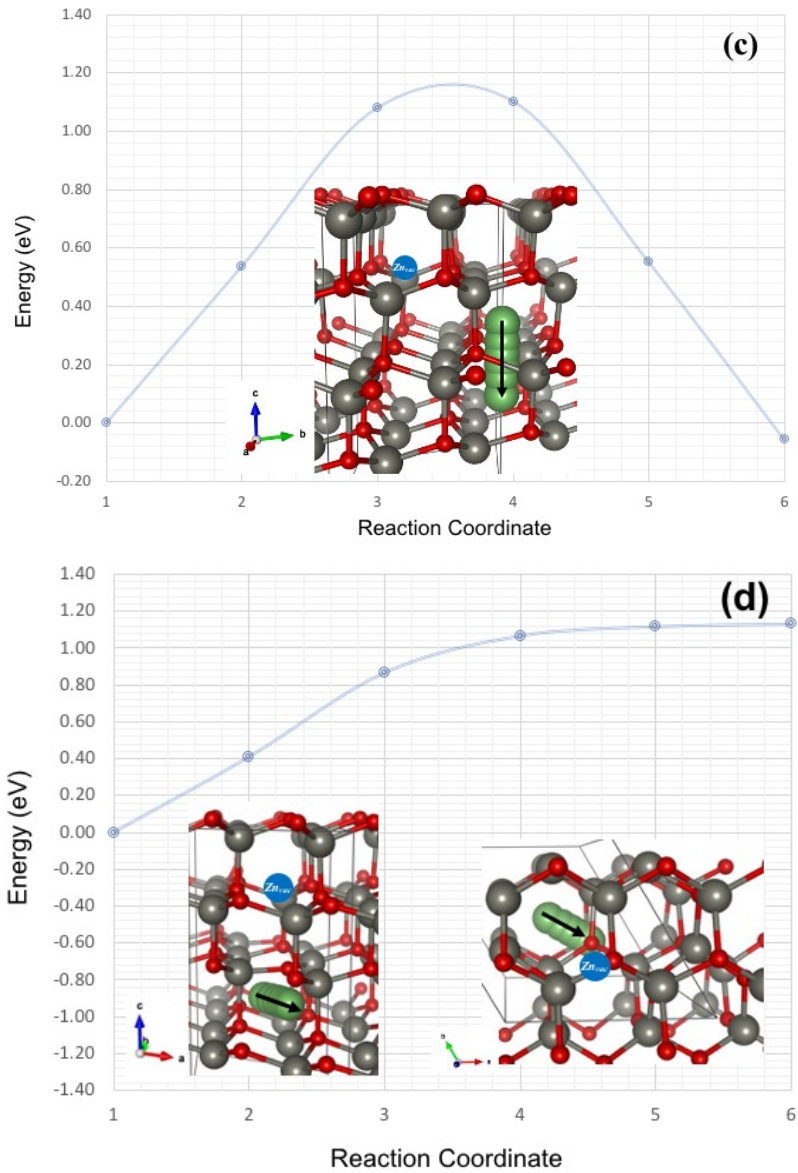

**Figure 3**. Li diffusion on ZnO with $Zn_{vac}$. (a) O-O path and (b) O-T Li diffusion paths at next-nearest octahedra relative from the $Zn_{vac}$. (c) O-O path and (d) O-T diffusion paths at next-next-nearest octahedra relative from $Zn_{vac}$ site. $Zn_{vac}$ is marked by the blue circle.

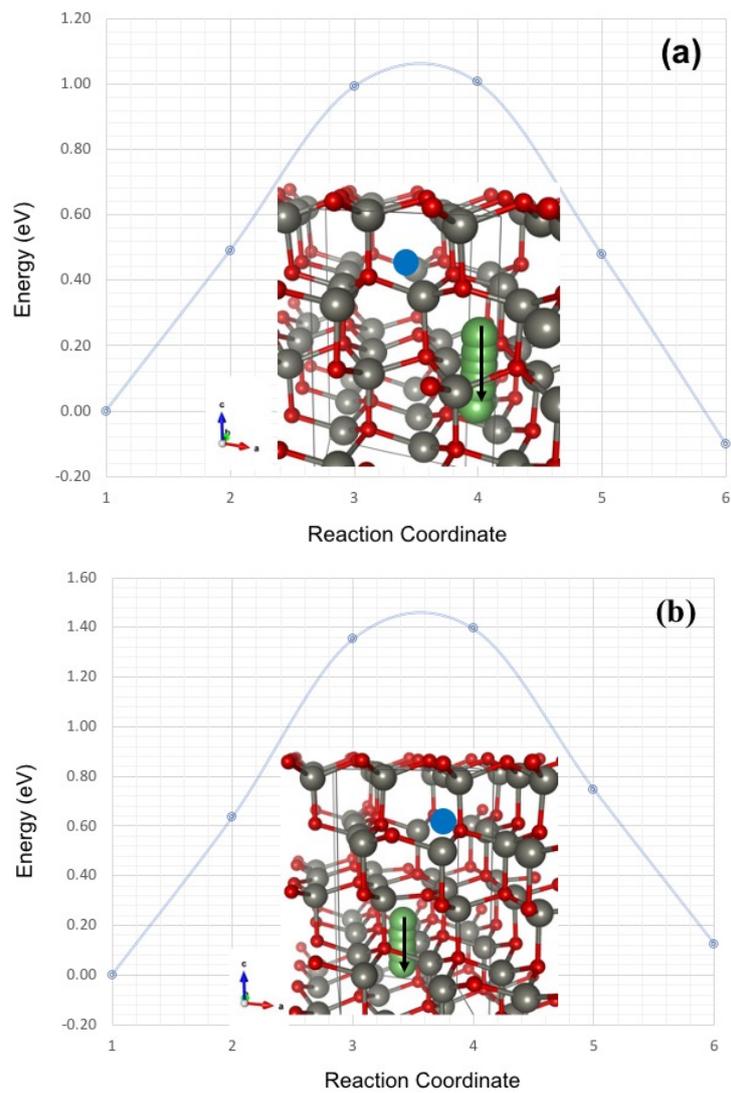

**Figure 4.** Li diffusion on ZnO with ZnO$_{vac\_pair}$. O-O path of Li diffusion on (a) next-nearest octahedra and (b) next-next-nearest octahedra relative from the ZnO$_{vac\_pair}$. ZnO$_{vac\_pair}$ is marked by the blue circle